\documentclass[11pt,twoside]{article}
\usepackage{asp2010}

\resetcounters

\begin{document}

\title{Cosmic star formation history and AGN evolution near and far: from AKARI to SPICA}
\author{Tomotsugu Goto$^1$, Takehiko Wada$^2$, Hideo Matsuhara$^2$, AKARI NEP team, AKARI all sky survey team, and SPICA MCS team
\affil{$^1$Dark Cosmology Centre, Niels Bohr Institute, Denmark\\
}
\affil{$^2$Institute of Space and Astronautical Science, JAXA, Japan}
}

\begin{abstract}
Infrared (IR) luminosity is fundamental to understanding the cosmic star formation history and AGN evolution, since their most intense stages are often obscured by dust. Japanese infrared satellite, AKARI, provided unique data sets to probe these both at low and high redshifts.
The AKARI performed an all sky survey in 6 IR bands (9, 18, 65, 90, 140, and 160$\mu$m) with 3-10 times better sensitivity than IRAS, covering the crucial far-IR wavelengths across the peak of the dust emission. Combined with a better spatial resolution, AKARI can measure the total infrared luminosity ($L_{TIR}$) of individual galaxies much more precisely, and thus, the total infrared luminosity density of the local Universe.
In the AKARI NEP deep field, we construct restframe 8$\mu$m, 12$\mu$m, and total infrared (TIR) luminosity functions (LFs) at 0.15$<z<$2.2 using 4128 infrared sources. A continuous filter coverage in the mid-IR wavelength (2.4, 3.2, 4.1, 7, 9, 11, 15, 18, and 24$\mu$m) by the AKARI satellite allows us to estimate restframe 8$\mu$m and 12$\mu$m luminosities without using a large extrapolation based on a SED fit, which was the largest uncertainty in previous work.
By combining these two results, we reveal dust-hidden cosmic star formation history and AGN evolution from $z$=0 to $z$=2.2, all probed by the AKARI satellite.
 The next generation space infrared telescope, SPICA, will revolutionize our view of the infrared Universe with superb sensitivity of the cooled 3m space telescope. We conclude with our survey proposal and future prospects with SPICA.
\end{abstract}

\section{Lessons from AKARI}

\subsection{Background}

Revealing the cosmic star formation history is one of the major goals of observational astronomy. However, UV/optical estimation only provides us with a lower limit of the star formation rate (SFR) due to obscuration by dust. 
A straightforward way to overcome this problem is to observe in the infrared, which can capture star formation activity invisible in the UV. 
The superb sensitivities of Spitzer and AKARI satellites have revolutionized the field.

In the local Universe, often used IR LFs are from the IRAS \citep[e.g.,][]{sanders2003,Goto2011IRAS} from 1980s, with only several hundred galaxies.
 In addition, bolometric infrared luminosities ($L_{IR,8-1000\mu m}$) of local galaxies were estimated using equation in P{\'e}rault (1987), which was a simple polynomial, obtained assuming a simple blackbody and dust emissivity. Furthermore, the reddest filter of IRAS was 100$\mu$m, which did not span the peak of the dust emission for most galaxies, leaving a great deal of uncertainty. 
Using deeper AKARI all sky survey data that cover up to 160$\mu$m,  we aim to measure local $L_{IR}$, and thereby the IR LF more accurately.

At higher redshifts, most of the Spitzer work relied on a large extrapolation from 24$\mu$m flux to estimate the 8, 12$\mu$m or total infrared (TIR) luminosity, due to the limited number of mid-IR filters.
AKARI has continuous filter coverage across the mid-IR wavelengths,  thus, allowing us to estimate mid-IR luminosity without using a large $k$-correction based on the SED models, eliminating the largest uncertainty in previous work. 
By taking advantage of this, we present the restframe 8, 12$\mu$m, and TIR LFs, and thereby the cosmic star formation history derived from these using the AKARI NEP-Deep data.\begin{figure}

\begin{center}
\includegraphics[scale=0.35]{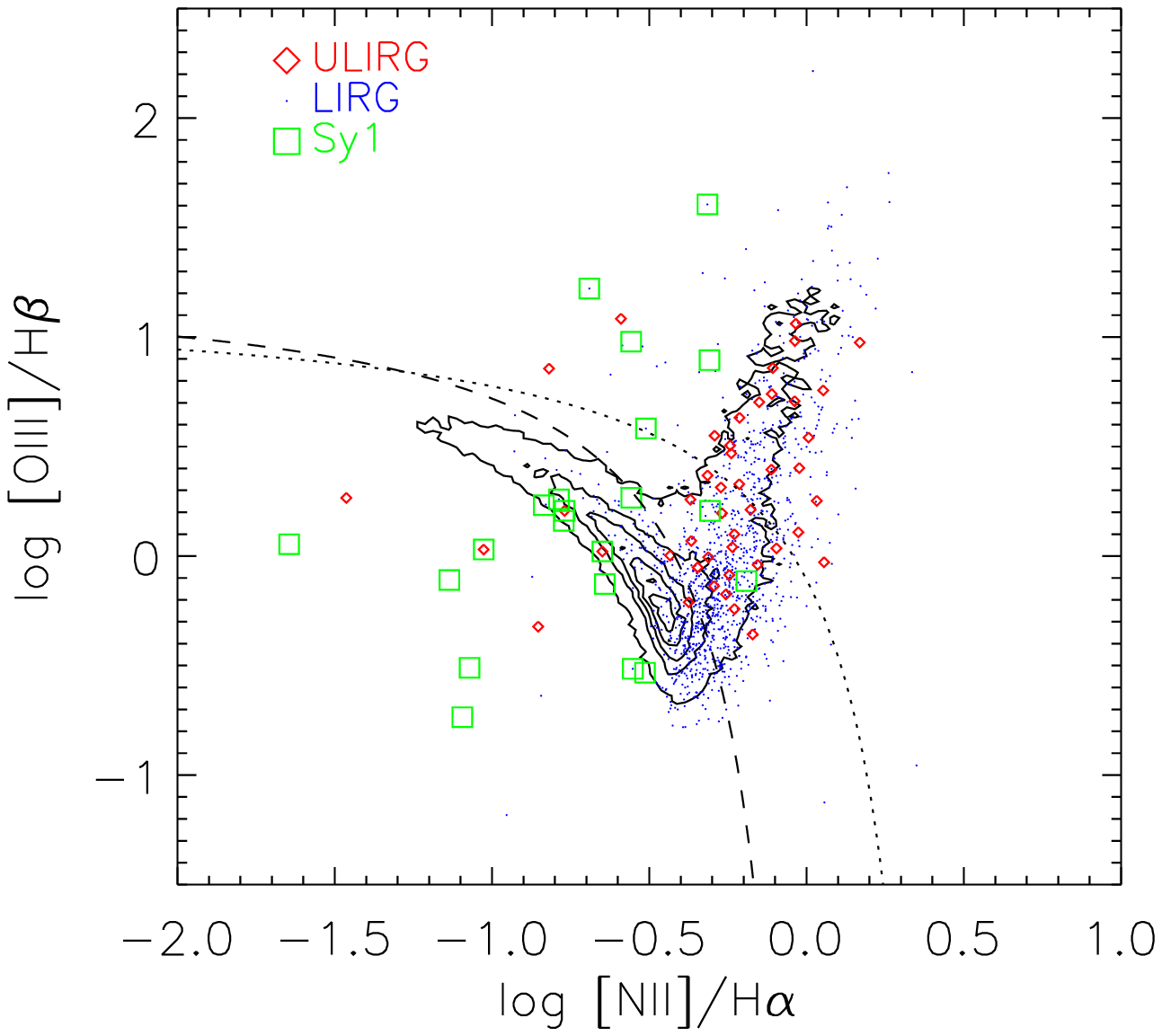}
\includegraphics[scale=0.35]{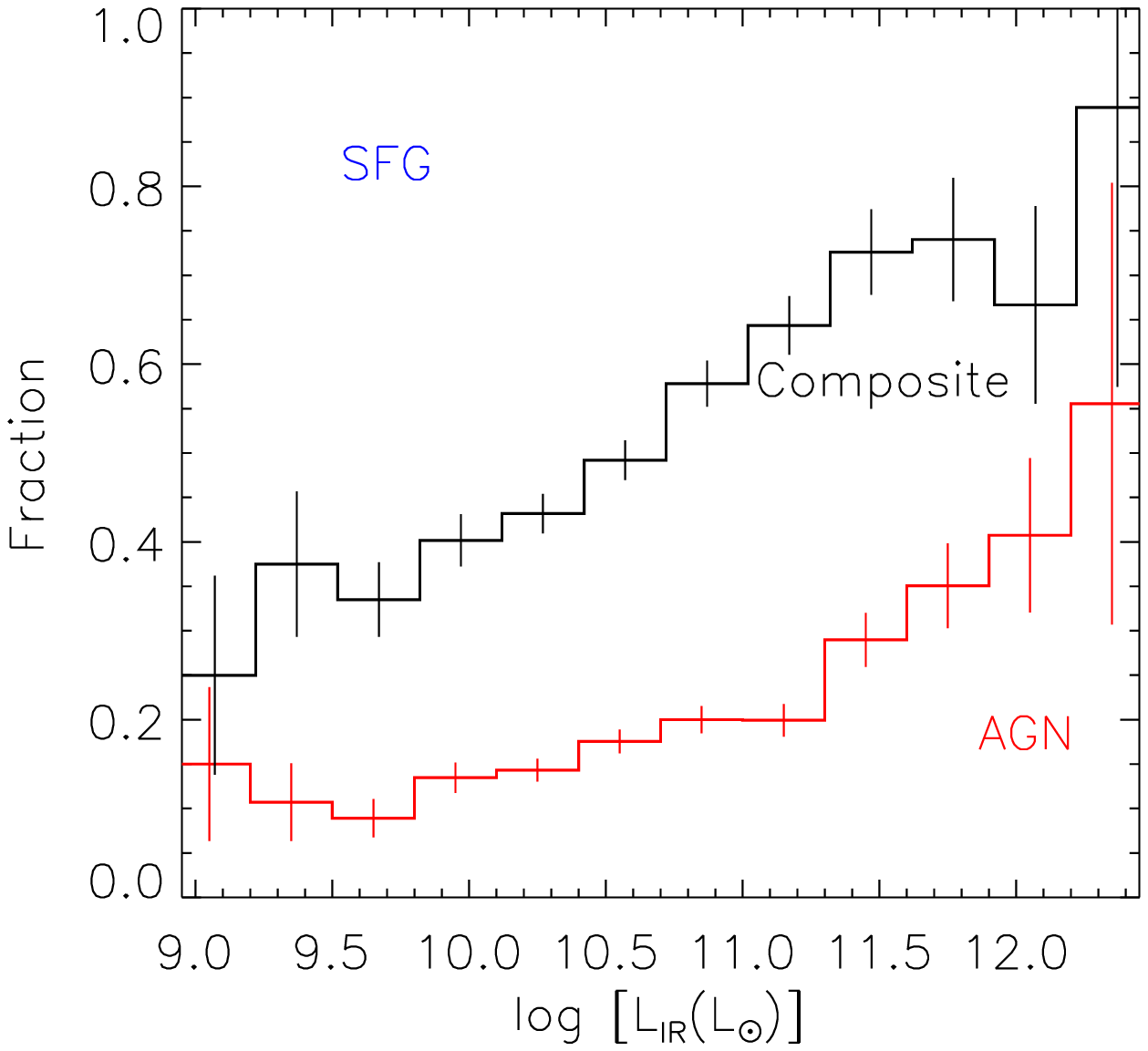}
\end{center}
\caption{
(left) Emission line ratios used to select AGNs from the AKARI all sky sample. The contour shows distribution of all galaxies in the SDSS with $r<17.77$ (regardless of IR detection). 
The dotted line is the criterion between starbursts and AGNs described in \citep{2001ApJ...556..121K}.
The dashed line is the criterion by \citep{2003MNRAS.346.1055K}.
Galaxies with line ratios higher than the dotted line are regarded as AGNs. 
Galaxies below the dashed line are regarded as star-forming. 
Galaxies between the dashed and dotted lines are regarded as composites. 
The blue and red dots are for ULIRGs, and LIRGs, respectively.
The green squares are Seyfert 1 galaxies identified by visual inspection of optical spectra. 
}\label{fig:BPT}
\caption{(right) Fractions of AGN and composite galaxies as  a function of $L_{IR}$.
AGN are classified using \citep{2001ApJ...556..121K} among galaxies with all 4 lines measured.  Composite galaxies include those classified as AGN using \citet{2003MNRAS.346.1055K}.
}\label{fig:AGN_fractions}
\end{figure}
%

\begin{figure}[!ht]
\begin{center}
\includegraphics[scale=0.35]{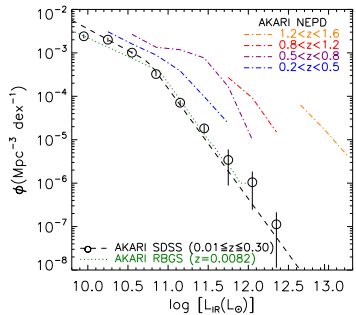}
\caption{
Infrared luminosity function of AKARI-SDSS galaxies. The $L_{IR}$ is measured using the AKARI 9, 18, 65, 90, 140 and 160$\mu$m fluxes through an SED fit. Errors are computed using 150 Monte Carlo simulations, added to a Poisson error.
The dotted lines show the best-fit double-power law. 
The green dotted lines show IR LF at $z$=0.0082 by \citep{Goto2011IRAS}.
The dashed-dotted lines are higher redshift results from the AKARI NEP deep field \citep{GotoKoyama2010,GotoTakagi2010}.
}\label{fig:LF}
\end{center}
\end{figure}

\subsection{{\bfseries AKARI All Sky Survey: low-$z$ Universe}}

\begin{figure*}
\includegraphics[scale=0.24]{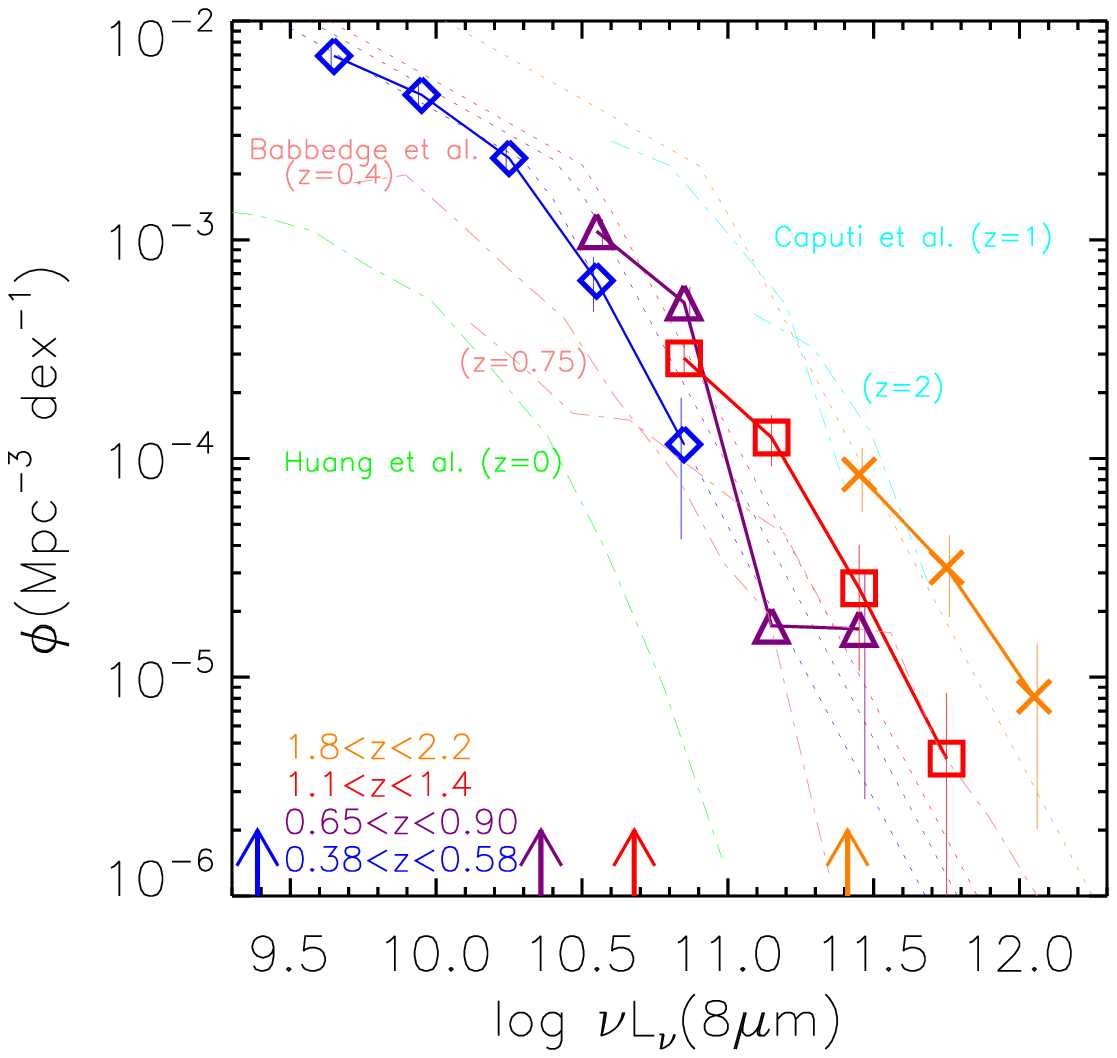}
\includegraphics[scale=0.24]{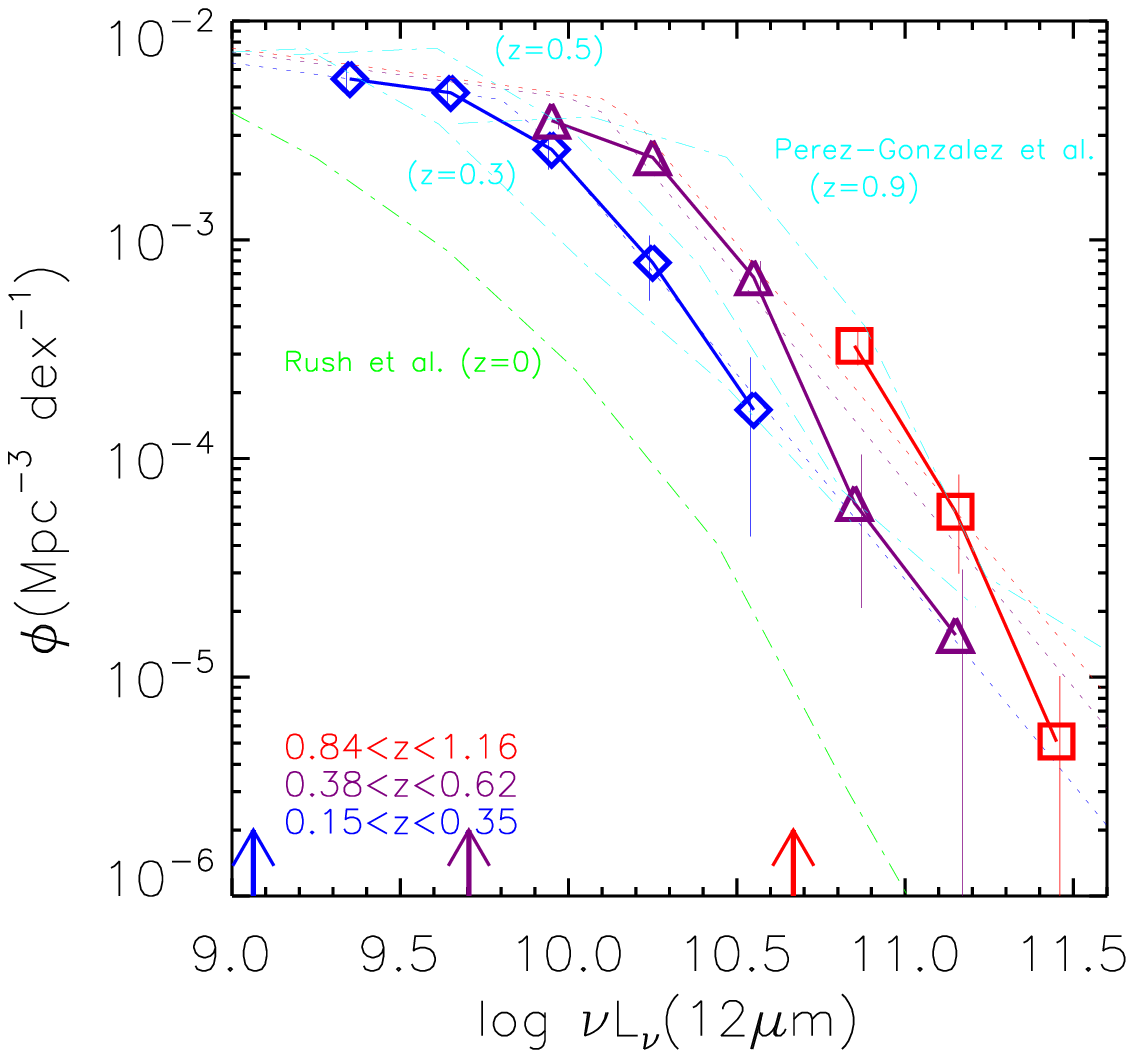}
\includegraphics[scale=0.24]{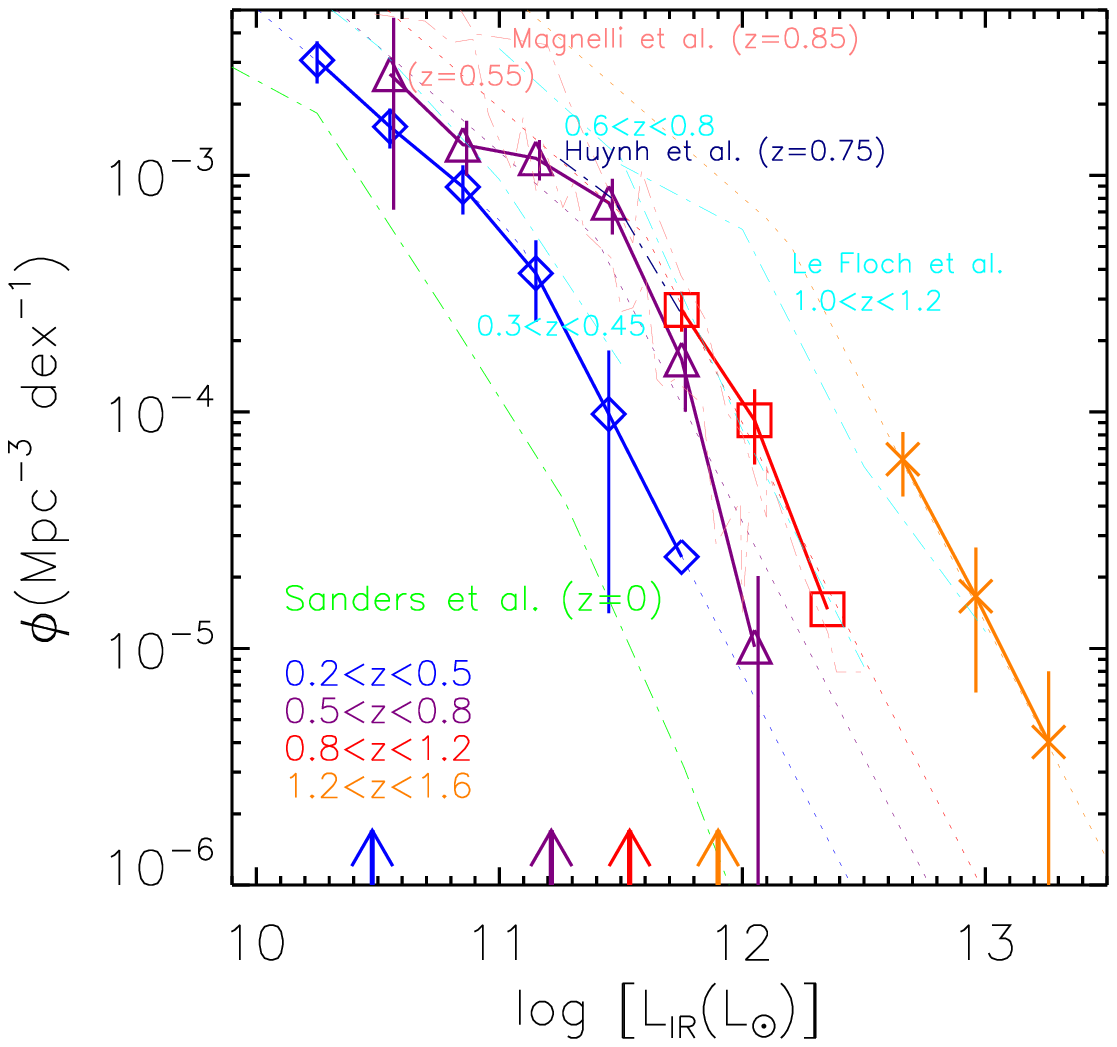}
\caption{
(left) Restframe  8$\mu$m LFs.
 The blue diamonds, purple triangles, red squares, and orange crosses show the 8$\mu$m LFs at $0.38<z<0.58, 0.65<z<0.90, 1.1<z<1.4$, and $1.8<z<2.2$, respectively. 
 The dotted lines show analytical fits with a double-power law.
 Vertical arrows show the 8$\mu$m luminosity corresponding to the flux limit at the central redshift in each redshift bin.
(middle)
Restframe  12$\mu$m LFs.
  The blue diamonds, purple triangles, and red squares show the 12$\mu$m LFs at $0.15<z<0.35, 0.38<z<0.62$, and $0.84<z<1.16$, respectively.
(right)
TIR LFs.  The redshift bins used are 0.2$<z<$0.5, 0.5$<z<$0.8, 0.8$<z<$1.2,  and 1.2$<z<$1.6. 
}\label{fig:8umlf}
\end{figure*}

\begin{figure}
\begin{center}
 \includegraphics[scale=.13]{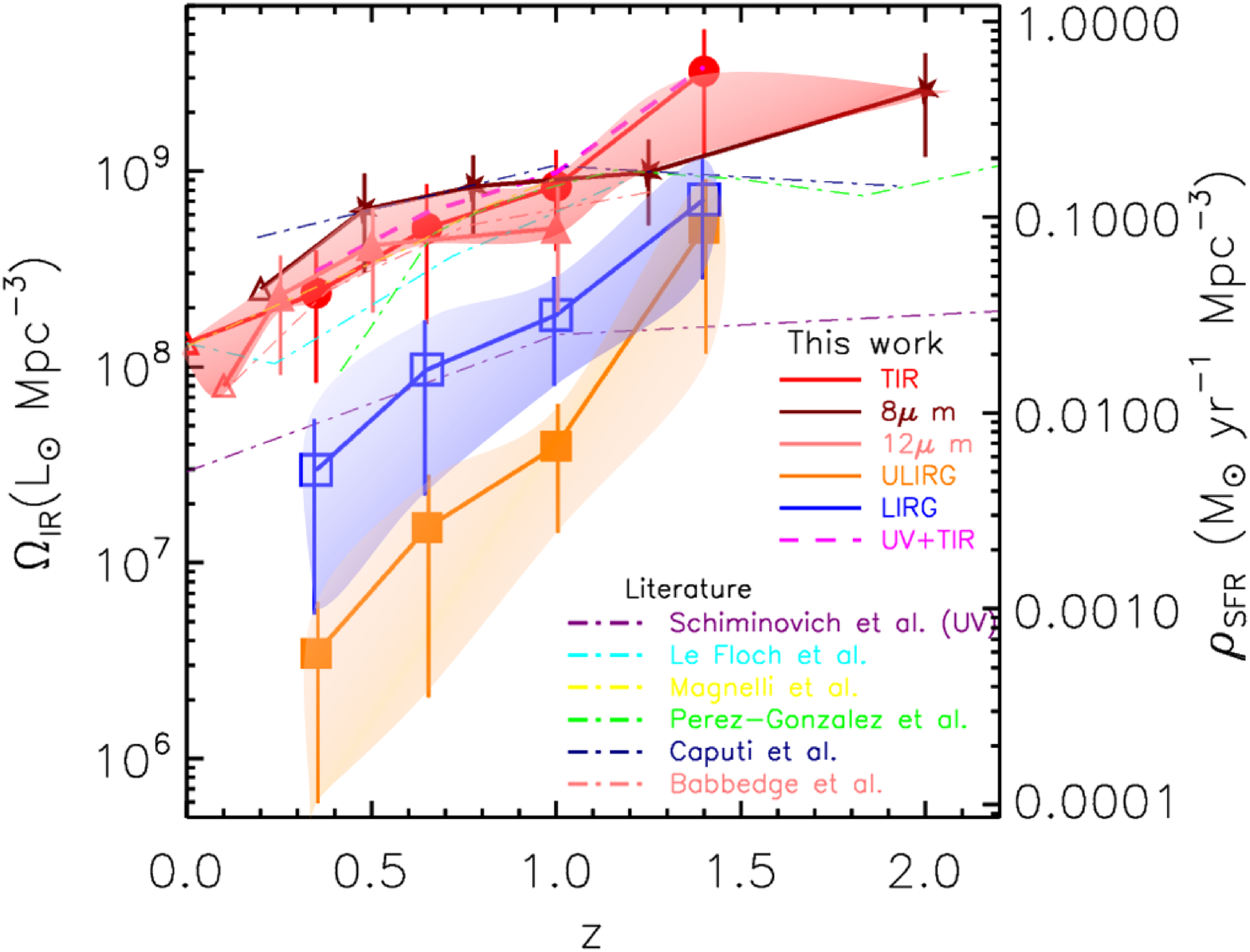}
 \includegraphics[scale=.09]{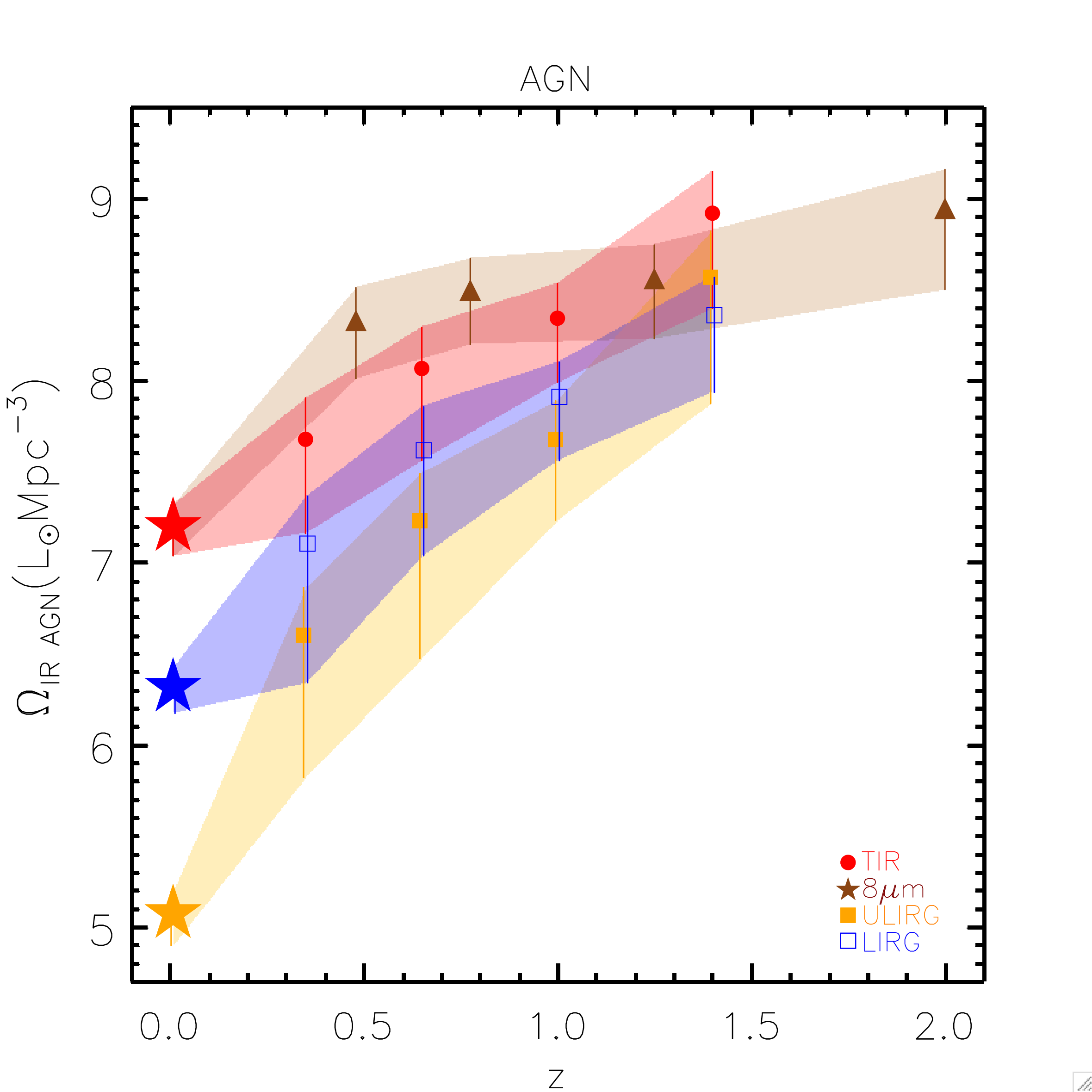}
\end{center}
\caption{(left)
Evolution of TIR luminosity density based on TIR LFs (red circles), 8$\mu$m LFs (stars), and 12$\mu$m LFs (filled triangles). The blue open squares and orange filled squares  are for LIRG and ULIRGs only, also based on our $L_{TIR}$ LFs.
}\label{fig:TLD_all}
\caption{
(right)
Evolution of TIR luminosity density by AGN.
 Results from the AKARI all sky survey is plotted with stars at $z$=0.0082. The red, blue and orange points show IR luminosity density from all AGN, from LIRG AGN only, and from ULIRG AGN only. 
Higher redshift results are from the AKARI NEP deep field \citep{GotoTakagi2010}, with contribution from star forming galaxies removed.
 Brown triangles are $\Omega_{IR}^{AGN}$ computed from the 8$\mu$m LFs \citep{GotoTakagi2010}.
}\label{fig:TLD_AGN_all}
\end{figure}

%

 AKARI performed an all-sky survey in two mid-infrared bands (centered on 9 and 18 $\mu$m) and in four far-infrared bands (65, 90, 140, and 160$\mu$m). 
 In addition to the much improved sensitivity and spatial resolution over its precursor (the IRAS all-sky survey), the presence of the 140 and 160$\mu$m bands is crucial to measure the peak of the dust emission in the FIR wavelength, and thus the $L_{IR}$ of galaxies.
We have cross-correlated the AKARI FIS bright source catalog with the SDSS DR7 galaxy catalog, obtaining 2357 cross-matched spectroscopic redshifts.

It is fundamental to separate IR contribution from two different physical processes; the star-formation and AGN activity.
In Fig. \ref{fig:BPT}, we use $[NII]/H\alpha$ vs $[OIII]/H\beta$ line ratios to classify galaxies into AGN or SFG (star-forming galaxies). It is interesting that the majority of (U)LIRGs are aligned along the AGN branch of the diagram, implying the 
AGN fraction is high among (U)LIRGs. This is more clearly seen in Fig. \ref{fig:AGN_fractions}, where we plot fractions of AGN as a function of $L_{IR}$.
These results agree with previous AGN fraction estimates \citep{2005MNRAS.360..322G}.
 Improvement in this work is that due to much larger statistics, we were able to show fractions of AGN in much finer luminosity bins, more accurately quantifying the increase. Especially, a sudden increase of $f_{AGN}$ at log$L_{IR}>$11.3 is notable due to the improved statistics in this work.

For these galaxies, we estimated total IR luminosities ($L_{IR}$) by fitting the AKARI photometry with SED templates. 
We used the {\ttfamily  LePhare} code\footnote{http://www.cfht.hawaii.edu/$^{\sim}$arnouts/lephare.html} to fit the infrared part ($>$7$\mu$m) of the SED. 
We fit our AKARI FIR photometry with the SED templates from Chary \& Elbaz (2001; CHEL hereafter), which showed most promising results among SED models tested by \citep{Goto2011IRAS}.

With accurately measured $L_{IR}$, we are ready to construct IR LFs.
Since our sample is flux-limited at $r=17.7$ and $S_{90\mu m}=0.7Jy$, we need to correct for a volume effect to compute LFs.
We used the 1/$V_{\max}$ method. 
We estimated errors on the LFs with 150 Monte Carlo simulations, added to a Poisson error.

In Fig. \ref{fig:LF}, we show infrared LF of the AKARI-SDSS galaxies.
The median redshift of our galaxies is $z$=0.031.

Once we measured the LF,  we can estimate the total infrared luminosity density by integrating the LF, weighted by the luminosity. We used the best-fit double-power law to integrate outside the luminosity range in which we have data, to obtain estimates of the total infrared luminosity density, $\Omega_{IR}$. Note that outside of the luminosity range we have data ($L_{IR}>10^{12.5}L_{\odot}$ or $L_{IR}<10^{9.8}L_{\odot}$), the LFs are merely an extrapolation and thus uncertain.

The resulting total luminosity density is  $\Omega_{IR}$= (3.8$^{+5.8}_{-1.2})\times 10^{8}$ $L_{\odot}$Mpc$^{-3}$.
Errors are estimated by varying the fit within 1$\sigma$ of uncertainty in LFs.
 Out of  $\Omega_{IR}$, 1.1$\pm0.1$\% is produced by LIRG ($L_{IR}>10^{11}L_{\odot}$), and only 0.03$\pm$0.01\% is by ULIRG ($L_{IR}>10^{12}L_{\odot}$). Although these fractions are larger than at $z$=0.0081 \citep[][]{Goto2011IRAS}, still a small fraction of  $\Omega_{IR}$ is produced by luminous infrared galaxies at $z$=0.031, in contrast to the high-redshift Universe.


\subsection{{\bfseries AKARI NEP Deep Field: high-$z$ Universe}}

The AKARI has observed the NEP deep field (0.4 deg$^2$) in 9 filters ($N2, N3, N4, S7, S9W$, $S11, L15, L18W$ and $L24$) to the depths of 14.2, 11.0, 8.0, 48, 58, 71, 117, 121 and 275 $\mu$Jy \citep[5$\sigma$,][]{2008PASJ...60S.517W}. 
This region is also observed in $BVRi'z'$ (Subaru),  $u'$ (CFHT), $FUV,NUV$ (GALEX), and $J,Ks$ (KPNO2m), with which we computed photo-z with $\frac{\Delta z}{1+z}$=0.043.  Objects which are better fit with a QSO template are removed from the analysis.
We used a total of 4128 IR sources down to 100 $\mu$Jy in the $L18$ filter.
We compute LFs using the 1/$V_{\max}$ method. Data are used to 5$\sigma$ with completeness correction. Errors of the LFs are from 1000 realization of Monte Carlo simulation.


\subsubsection{{\bfseries Restframe 8$\mu$m luminosity functions}}

Monochromatic 8$\mu$m luminosity ($L_{8\mu m}$) is known to correlate well with the TIR luminosity \citep[][]{Goto2011SDSS}), especially for star-forming galaxies, because the rest-frame 8$\mu$m flux is dominated by prominent PAH features such as at 6.2, 7.7 and 8.6 $\mu$m.
 The left panel of Fig. \ref{fig:8umlf} shows a strong evolution of 8$\mu$m LFs.
 Overplotted previous work had to rely on SED models to estimate $L_{8\mu m}$ from the Spitzer $S_{24\mu m}$ in the MIR wavelengths where SED modeling is difficult due to the complicated PAH emissions. Here, AKARI's mid-IR bands are advantageous in  directly observing redshifted restframe 8$\mu$m flux in one of the AKARI's filters, leading to more reliable measurement of 8-$\mu$m LFs without uncertainty from the SED modeling. 

\subsubsection{{\bfseries Restframe 12$\mu$m luminosity functions}}

 12$\mu$m luminosity ($L_{12\mu m}$) represents mid-IR continuum, and known to correlate closely with TIR luminosity \citep{perez2005}. 
  The middle panel of Fig. \ref{fig:8umlf} shows a strong evoltuion of 12$\mu$m LFs.
 Here the agreement with previous work is better because (i)  12$\mu$m continuum is easier to be modeled, and (ii) the Spitzer also captures restframe 12$\mu$m in $S_{24\mu m}$ at $z$=1.

\subsubsection{{\bfseries Total infrared luminosity functions}}
 Lastly, we show the TIR LFs in the right panel of Fig. \ref{fig:8umlf}. 
We used \citet{Lagache2003}'s SED templates to fit the photometry using the AKARI bands at $>$6$\mu$m ($S7,S9W$, $S11,L15,L18W$ and $L24$). 
 The TIR LFs show a strong evolution compared to local LFs. 
 At $0.25<z<1.3$, $L^*_{TIR}$ evolves as $\propto (1+z)^{4.1\pm0.4}$.


\subsubsection{{\bfseries Cosmic star formation history}}

We fit LFs in Fig. \ref{fig:8umlf} with a double-power law, then integrate to estimate total infrared luminosity density at various $z$. The restframe 8 and 12$\mu$m LFs are converted to $L_{TIR}$ using \citet{perez2005,caputi2007}  before integration.
 The resulting evolution of the TIR density is shown in Fig. \ref{fig:TLD_all}.
 The right axis shows the star formation density assuming Kennicutt (1998). 
 We obtain $\Omega_{IR}^{SFG}(z) \propto (1+z)^{4.1\pm 0.4}$.
Comparison to $\Omega_{UV}$ by \citet{Schiminovich2005} suggests that $\Omega_{TIR}$ explains 70\% of $\Omega_{total}$ at $z$=0.25, and that by $z$=1.3, 90\% of the cosmic SFD is explained by the infrared. This implies that $\Omega_{TIR}$ provides good approximation of the  $\Omega_{total}$ at $z>1$.
 
 In Fig. \ref{fig:TLD_all}, we also show the contributions to $\Omega_{TIR}$ from LIRGs and ULIRGs.
 From $z$=0.35 to $z$=1.4, $\Omega_{IR}$ by LIRGs increases by a factor of $\sim$1.6, and 
  $\Omega_{IR}$ by ULIRGs increases by a factor of $\sim$10.
More details are in Goto et al. (2010a).

\subsubsection{{\bfseries Cosmic AGN accretion history}}

We have separated the  $\Omega_{IR}^{SFG}$ from  $\Omega_{IR}^{AGN}$. 
Therefore, we can also investigate $\Omega_{IR}^{AGN}$.
By integrating IR LF$_{AGN}$, we show the evolution of $\Omega_{IR}^{AGN}$ in Fig. \ref{fig:TLD_AGN_all}, which shows a strong evolution with increasing redshift. 
 At a first glance, both $\Omega_{IR}^{AGN}$ and $\Omega_{IR}^{SFG}$ show rapid evolution, suggesting that the correlation between star formation and black hole accretion rate continues to hold at higher redshifts, i.e., galaxies and black holes seem to be evolving hand in hand.
 When we fit the evolution with (1+z)$^{\gamma}$, we find
 $\Omega_{IR}^{AGN}\propto$(1+z)$^{4.1\pm0.5}$.
 A caveat, however, is that $\Omega_{IR}^{AGN}$ estimated in this work is likely to include IR emission from host galaxies of AGN, although in optical the AGN component dominates. Therefore, the final conclusion must be drawn from a multi-component fit based on better sampling in FIR by Herschel or SPICA, to separate AGN/SFG contribution to $L_{IR}$.
The contribution by ULIRGs quickly increases toward higher redshift;  By $z$=1.5, it exceeds that from LIRGs. Indeed, we found 
$\Omega_{IR}^{AGN}(ULIRG)\propto$(1+z)$^{8.7\pm0.6}$ and 
$\Omega_{IR}^{AGN}(LIRG)\propto$(1+z)$^{5.4\pm0.5}$.


%
%
%

\section{Wide area survey with SPICA's Mid-infrared Camera and Spectrograph (MCS)}

\begin{figure}[!tpb]
   \begin{center}
\includegraphics[width=4cm]{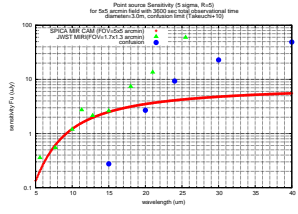}
   \end{center}
   \caption{ \label{fig:sensitivity_confusion}
  Comparison of survey speed for 5$\times$5 arcmin$^2$ area (1 hour, S/N=5, R=5). 
  SPICA's confusion limit is shown in the blue filled-circles.  }  
\end{figure}

SPICA \citep[][]{Nakagawa_2011} 
is the next-generation, space infrared (5-210$\mu$m) telescope (target launch in 2022). 
 With its 3.2-meter telescope cryogenically cooled to 6 Kelvin, SPICA is 100 times more sensitive than its precursors. 
 Its Mid-infrared Camera and Spectrograph \citep[MCS;][]{KatazaMCS2012,WadaMCS2012} has a large field of view of 5$\times$5 arcmin, and is sensitive in 5-38$\mu$m.
In Figure \ref{fig:sensitivity_confusion}, we compare SPICA/MCS's survey speed (in 5$\times$5 arcmin$^2$ area) with that of James Webb Space Telescope (JWST).  The figure shows that SPICA's survey speed is comparable to JWST at $<15\mu$m, and superior at $>18\mu$m. Note however at $>20\mu$m, SPICA reaches galaxy confusion limit (the blue filled circles) in less than one hour.
Taking advantage of the wide-field of view, and superb sensitivity of the MCS, we propose a wide area confusion-limited imaging survey using all the broad filters of MCS.

\subsection{Survey design}\label{surveys}
\subsubsection{Deep survey}
       It has been known that infrared (IR) properties of galaxies depend on galaxy environment. For example, \citet{2005MNRAS.360..322G} showed that galaxy density distribution depends on IR luminosity, in a sense that more IR luminous galaxies are in less dense environment in the local Universe.  \citet{GotoKoyama2010} showed that the shape of the  restframe 8$\mu$m LFs depends on the local galaxy density. 
       \citet{2008MNRAS.391.1758K}  found the excess of 15-$\mu$m-detected galaxies in the medium-density environments.
       These examples show that we need to investigate various environments from dense cluster cores to rarefied field, to fully understand infrared properties of galaxies and their evolution. 
 As a minimum of such study field, we propose one deg$^2$ survey area. Such a field covers 30 by 30 Mpc at z=3, including one massive cluster in the survey area, allowing us to investigate from dense cluster core to the rarefied field continuously. 

 This survey is also where the MCS is technically advantageous;  the MCS's field of view (5 by 5') is much larger than that of the JSWT/MIRI (1.7 by 1.3'). Therefore, in terms of the survey speed of 5$\times$5' area or larger, the MCS is comparably sensitive to the JWST/MIRI even at  $<$20$\mu$m (Fig. \ref{fig:sensitivity_confusion}). At longer wavelength of $>$20$\mu$m, the MCS is more sensitive than the JWST/MIRI. 

We need the depth of $\sim$1 $\mu$Jy at 10 $\mu$m to investigate ULIRGs up to z=3 .
 In order to fully sample mid-IR SED, we aim to image the area in all 8 broad filters of WFC-S channel and all 7 broad filters of WFC-L channel. Since in the WFC-L wavelength range, we reach the galaxy confusion limit in a short exposure time ($\sim$15 min), and
because the WFC-S and WFC-L channel can observe simultaneously,  the imaging in the 7 filters of the WFC-L channel can be obtained during the longer exposure of the WFC-S channel. In total, we need  9.67 hours per FoV (net exposure time of 1 hour per filter).
   We cover 0.9 deg$^2$ by 12$\times$12 tiling. This will take 1392 hours, or 60 days. This is 6\% of 2.5 years of SPICA mission lifetime.

The survey needs to be coordinated to be in the same region as the far-infrared SAFARI survey. The SAFARI survey needs $\sim$1000 hours to observe 1 deg$^2$ region, to the depth of z=3 ULIRGs. Therefore, in total, we need $\sim$2600 hours, or 3.6 months of SPICA time to observe the field. This is a large amount (12\%) of the SPICA mission lifetime, and therefore, the field needs to be of a good visibility. Ancillary data from 
other wavelengths are desirable. We propose the north ecliptic pole or south ecliptic pole regions as candidates. IR cirrus confusion needs to be smaller than galaxy confusion. We need to avoid regions with $>$3 MJy/sr. 

The near-infrared FPC camera (5$\times$5' field of view, 5 broad filters in 0.7-5 $\mu$m) can observe while the MCS is taking image (one of the HRS or MRS cameras need to be off). Therefore, in this region, near-IR image in 5 broad filters of FPC can be obtained simultaneously. The depth in one hour FPC exposure is expected to detect ULIRGs at z=3. 
These near-IR photometry will be useful to compute photometric redshift of MCS sources combined with the ground-based optical imaging. 

The WFC-L will expose for one hour in each filter, while it reaches confusion limit in 15 min. Therefore, $>$4 times redundant data will be obtained in each filter of WFC-L. The survey needs to be carefully planned to sample variability of the sources in various time scales. 

The survey volume at z=3 is 1.5$\times$10$^5$ Mpc$^3$ (with $\Delta$z of 0.2). 
The number density of ULIRGs at z=3 is 10$^{-3}$ Mpc$^{-3}$ dex$^{-1}$.
 Therefore, we expect to detect 150 ULIRGs at z=3 in a slice of $\Delta$z of 0.2.

At z=1, the survey volume is 4.0$\times$10$^5$ Mpc$^3$ (with $\Delta$z of 0.2). 
The number density of ULIRGs measured by AKARI at z=1 is 10$^{-4}$ Mpc$^{-3}$ dex$^{-1}$ \citep[Fig.\ref{fig:TLD_all};][]{GotoTakagi2010}.
Therefore, we expect to detect 40 ULIRGs at z=1 in a slice of $\Delta$z of 0.2.

\subsubsection{Wide survey}
 Complementary to the deep survey, the wide survey aims to detect rare objects that could not be found in numbers in the deep survey. For example, the number density of hyper luminous infrared galaxies at z=3 is 10$^{-4}$ Mpc$^{-3}$ dex$^{-1}$. In the volume of deep survey (1.5$\times$10$^5$Mpc$^3$, $\Delta$z of 0.2 at z=3), there will be only 15 hyper LIRGs, and thus, not enough for statistical study.
 In the wide survey, we aim to survey 10 deg$^2$ to the limit of 3-30 $\mu$Jy. This will give us a 10 times larger volume, and thus, we will create a large enough sample for statistical study with 150 hyper-LIRGs.

To reach 3 $\mu$Jy,  the total time requred is 0.975 hrs per FoV.
38 by 38 tiling can cover 9.05 deg$^2$.
Total telescope time required is then 1407 hours, or 6\% of 2.5 years of mission lifetime.

It is ideal that central part of the wide field to be covered by the deep survey, so that we understand the location of the deep field in terms of larger scale. However, both surveys require significant fraction (6\% each) of mission lifetime, and both require SAFARI counterpart surveys and various follow-up spectroscopic observation. Considering these factors, the deep and wide surveys are likely to be separated one at NEP and another near SEP.


%

\bibliography{tgoto}

\end{document}